\documentclass[twocolumn,prb,showpacs,citeautoscript,floatfix]{revtex4}
\usepackage{array}
\usepackage{multirow}
\usepackage{amssymb}
\usepackage{graphicx}
\input{epsf}

\begin{document}

\title{Suppression of the Charge-Density-Wave State in Sr$_{10}$Ca$_{4}$Cu$_{24}$O$_{41}$ by External Pressure}
\author{C.~A.~Kuntscher$^{1,*}$, A. Huber$^{1}$, and M. H\"ucker$^{2}$}
\affiliation{$^{1}$Experimentalphysik~2,~Universit\"at~Augsburg,~
D-86195~Augsburg,~Germany }
\affiliation{$^{2}$Brookhaven National Laboratory, Upton, NY 11973-5000, USA}

\date{\today}

\begin{abstract}
The influence of external pressure on the charge-density-wave (CDW) ground state of the quasi-one-dimensional two-leg ladder
compound Sr$_{10}$Ca$_{4}$Cu$_{24}$O$_{41}$ has been studied by optical reflectivity measurements as a function of temperature (10 - 300~K) and pressure $P$ (0.3 - 4.3~GPa) over the spectral range 580 - 6000 cm$^1$. With increasing pressure the CDW transition temperature $T_{CDW}$ decreases with the linear pressure coefficient $\approx$-70~K/GPa, and above $\approx$3~GPa the CDW phase is suppressed at all temperatures. This behavior is similar to that in compounds Sr$_{14-x}$Ca$_x$Cu$_{24}$O$_{41}$ with increasing Ca content $x$ at ambient pressure, with the simple scaling
$x \approx 3\cdot P(GPa)$. The size of the CDW gap decreases with increasing pressure, whereas the dimensionality of the high-temperature insulating phase in Sr$_{10}$Ca$_{4}$Cu$_{24}$O$_{41}$ within the ladder plane
is hardly affected by external pressure.
\end{abstract}


\maketitle

\section{Introduction}

The cuprate spin ladder compounds have been studied extensively experimentally as well as theoretically
during the last years, since they possess many similarities with the two-dimensional cuprate high-temperature
superconductors and therefore provide a playground to explore the interesting physics.\cite{Dagotto96,Vuletic06}
Furthermore, compounds based on spin ladders exhibit novel properties - such as spin-charge separation,
spin-gapped metallic state, and superconductivity competing with an insulating hole Wigner crystal
phase (often simply called charge-density-wave phase) - which justify extensive studies.

The quasi-one-dimensional compounds $M_{14}$(CuO$_2$)$_{10}$(Cu$_2$O$_3$)$_7$ with $M$=La,Y,Sr or Ca comprise Cu$_2$O$_3$ planes with two-leg
ladders alternating along the $b$ axis with planes containing CuO$_2$ chains with edge-shared CuO$_4$ plaquettes.
They belong to the class of low-dimensional transition-metal oxides with strong electronic correlations.
The two-leg ladder subsystem can be viewed as rung singlets. It has been predicted that ladder compounds can be doped with holes, which might lead to a superconducting (SC) state \cite{Dagotto96}.
The current picture is that the doped
holes are distributed such that a ladder rung is either doubly occupied by holes or empty, and the
hole doping can be achieved by Ca doping or by applying pressure.
The formation of charged hole pairs upon doping can give rise to either a SC state or a charge-density-wave (CDW) state. Indeed, superconducting properties below T$_C$=12 K were observed for the Ca-doped compound Sr$_{0.4}$Ca$_{13.6}$Cu$_{24}$O$_{41.8}$ at a pressure of 3~GPa.\cite{Uehara96} The delicate balance between the two different ground states (SC, CDW) is still not understood.
Their competition is suggested to be similar to the competition believed to occur between the ordered
stripes and superconductivity in the two-dimensional cuprate systems \cite{Tranquada97},
which makes the spin ladder an important reference system for the overall understanding of
the copper oxide materials.

The compound La$_6$Ca$_8$Cu$_{24}$O$_{41}$ is a Mott insulator, with an energy gap of
approximately 2~eV in size, which can be doped with holes by substitution of La by Sr, as in
La$_{6-y}$Sr$_y$Ca$_8$Cu$_{24}$O$_{41}$.
For the complete substitution of La by Sr the maximum hole doping of n$_h$=6 holes per formula unit is reached
and the average copper valence is +2.25.
The resulting material Sr$_6$Ca$_8$Cu$_{24}$O$_{41}$ belongs to the compounds Sr$_{14-x}$Ca$_x$Cu$_{24}$O$_{41}$,
which constitute the most heavily studied spin ladder systems, since they exhibit superconductivity for high Ca content.
The distribution of the holes among the ladder and chain subsystems as a function of Ca doping,
temperature, and pressure is one important issue in this material, since it determines its physical properties.
Contradictory experimental results for the hole distribution have been found with different experimental techniques, among them x-ray absorption and optical spectroscopy.\cite{Osafune97,Nucker00,Piskunov05,Rusydi07,Kabasawa08,Ma09,Deng11,Ilakovac12,Huang13}

In Sr$_{14}$Cu$_{24}$O$_{41}$ the doped holes in the ladder subsystem form an unconventional CDW ground state at low temperatures: Dc resistivity, low-frequency dielectric, and optical spectroscopy,
revealed an insulator-to-insulator transition during cooling down related to the development of
a CDW \cite{Gorshunov02,Blumberg02,Vuletic03,Vuletic05,Abbamonte04}.
According to the frequency-dependent complex dielectric response Ca doping suppresses the CDW state:
The transition temperature $T_{CDW}$ decreases from 210~K for $x$=0 to 7~K for $x$=9.\cite{Vuletic03} The size of the CDW energy gap $\Delta_{CDW}$ was determined from
the suppression of spectral weight below $\Delta_{CDW}$ in the optical conductivity spectra for the polarization
{\bf E} along the most conducting direction ($c$ axis). It decreases from $\Delta_{CDW}$=112~meV $x$=0 to 2.5~meV for $x$=9.\cite{Vuletic03} In contrast to these findings, resonant x-ray scattering results evidenced the existence of charge order only in compounds with 10$\leq$$x$$\leq$12, in addition to the $x$=0 compound.\cite{Rusydi06} The existence of a CDW for the Ca content $x$=11 is supported by an optical study.\cite{Osafune97a}

Obviously, the ground state of the Sr$_{14}$Cu$_{24}$O$_{41}$ system is very sensitive regarding Ca doping and external pressure. The substitution of Sr by Ca leads to a chemical pressure effect in the system due to the smaller ionic radius of the latter. According to x-ray diffraction experiments the main effect of both chemical and external pressure is a strong decrease in the lattice parameter $b$ and hence a reduced distance between the ladder and chain layers increasing the interlayer coupling.\cite{McCarron88,Isobe05,Pachot05}
In a first attempt, the SC and CDW ground states have been included in a
three-dimensional phase diagram with temperature, Ca content $x$, and pressure as the three variables.\cite{Vuletic06,Hess01,Vuletic03,Vuletic05,Uehara96,Motoyama02,Nagata98,Isobe98}
However, this phase diagram can only be considered as preliminary, since not all combinations of the three variables
have been realized experimentally up to now. In this paper we focus on the influence of external pressure on the CDW ground state in Sr$_{10}$Ca$_4$Cu$_{24}$O$_{41}$, which we study by optical reflectivity measurements as a function of temperature and pressure.

\begin{figure}[t]
\includegraphics[scale=0.43]{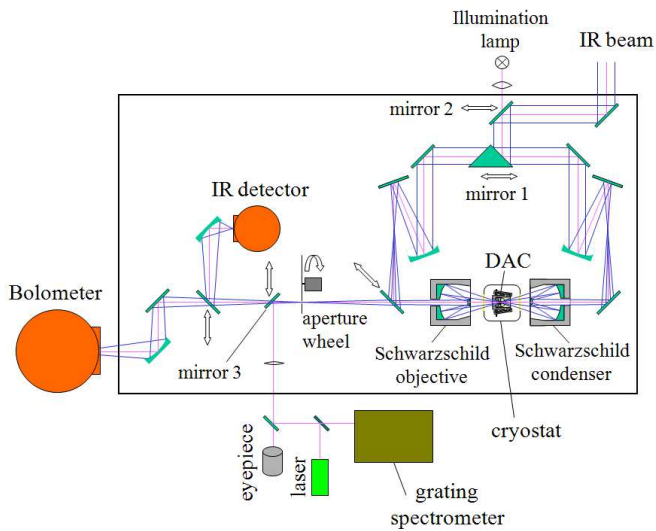}
\caption{Scheme of the home-built infrared microscope coupled to a Bruker Vertex 80/v FTIR spectrometer for
high-pressure and low-temperature optical reflectivity and transmission measurements.
Depending on the position of mirror 1 the setup can be operated in either reflection or transmission mode.
The radiation is focused on the sample by Schwarzschild objectives.
For the visible inspection of the sample mirror 2 is moved, so that the light from the illumination lamp is passing to the sample, and additionally mirror 3 is moved to guide the light towards the eyepiece.
For the {\it in-situ} pressure determination the ruby balls in the DAC are excited by a green laser ($\lambda$=532~nm), and the ruby fluorescence spectra are recorded by a grating spectrometer.
For polarization-dependent measurements a polarizer is added to the optical scheme.}
\label{fig:lthp}
\end{figure}

\begin{figure}
\includegraphics[width=0.8\columnwidth]{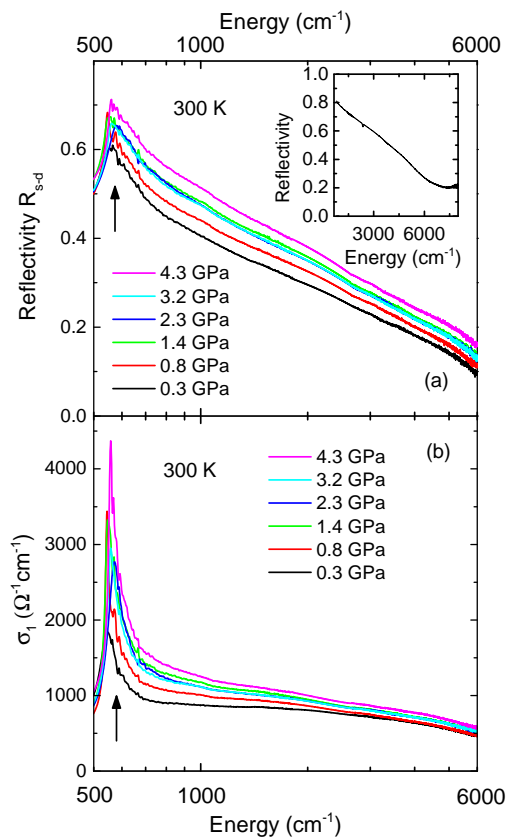}
\caption{(a) Room-temperature reflectivity spectra R$_{s-d}$ and corresponding (b) optical conductivity spectra of Sr$_{10}$Ca$_{4}$Cu$_{24}$O$_{41}$ for {\bf E}$\parallel$$c$ for various applied pressures. The arrows mark the phonon mode region. The inset in (a) depicts the mid-infrared reflectivity spectrum of the free-standing sample, taken outside the DAC, for illustrating the plasma edge.}
\label{fig:room-temp}
\end{figure}

\begin{figure*}
\includegraphics[width=10.5cm]{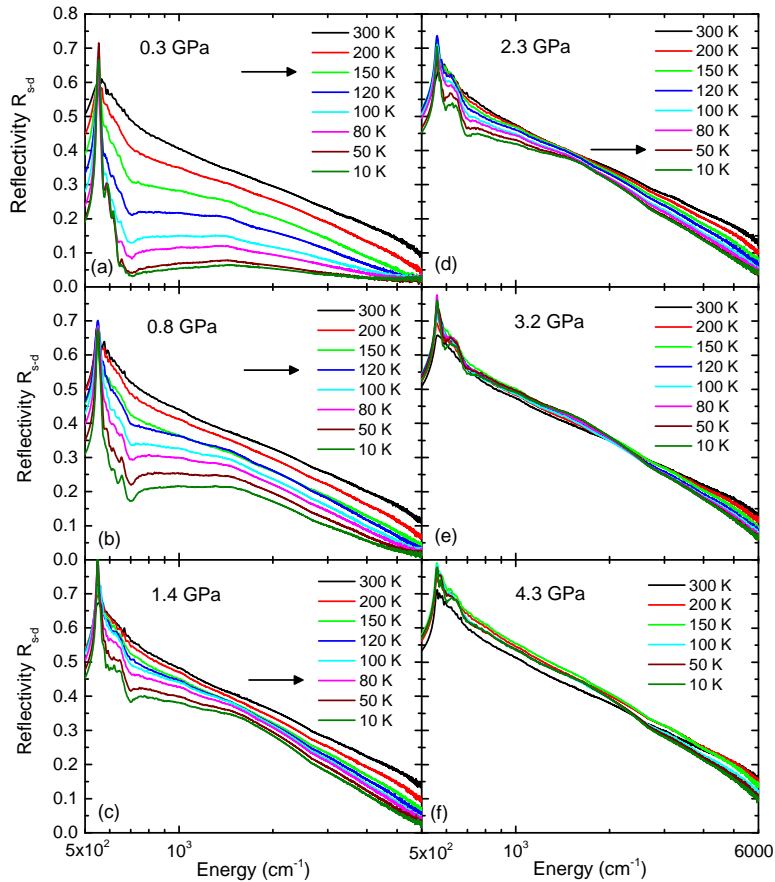}
\caption{Reflectivity spectra R$_{s-d}$ of Sr$_{10}$Ca$_{4}$Cu$_{24}$O$_{41}$ for {\bf E}$\parallel$$c$ as a function of temperature for various pressures. The arrows mark the transition temperature T$_{CDW}$, as determined from the change in slope in the reflectivity spectra.}
\label{fig:refl}
\end{figure*}

\begin{figure}
\includegraphics[width=0.8\columnwidth]{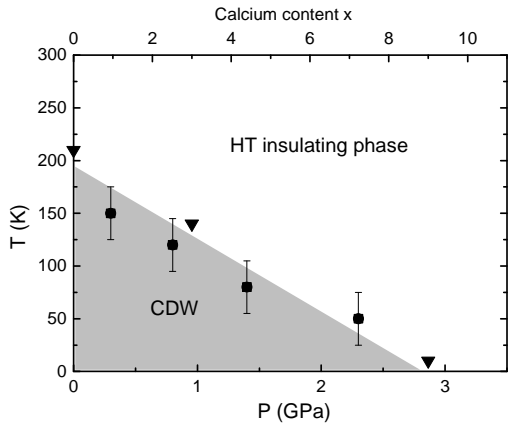}
\caption{Temperature-pressure phase diagram of Sr$_{10}$Ca$_{4}$Cu$_{24}$O$_{41}$ showing the high-temperature (HT)  insulating phase and the charge-density-wave (CDW) phase.
The filled squares mark the CDW transition temperature $T_{CDW}$ as extracted from the pressure- and temperature-dependent reflectivity spectra.
The filled triangles mark $T_{CDW}$ for ambient-pressure Sr$_{14-x}$Ca$_{x}$Cu$_{24}$O$_{41}$ as a function of Calcium content $x$ (upper horizontal axis) from Ref.\ \onlinecite{Vuletic03}.} \label{fig:pt}
\end{figure}

\begin{figure*}
\includegraphics[width=11.5cm]{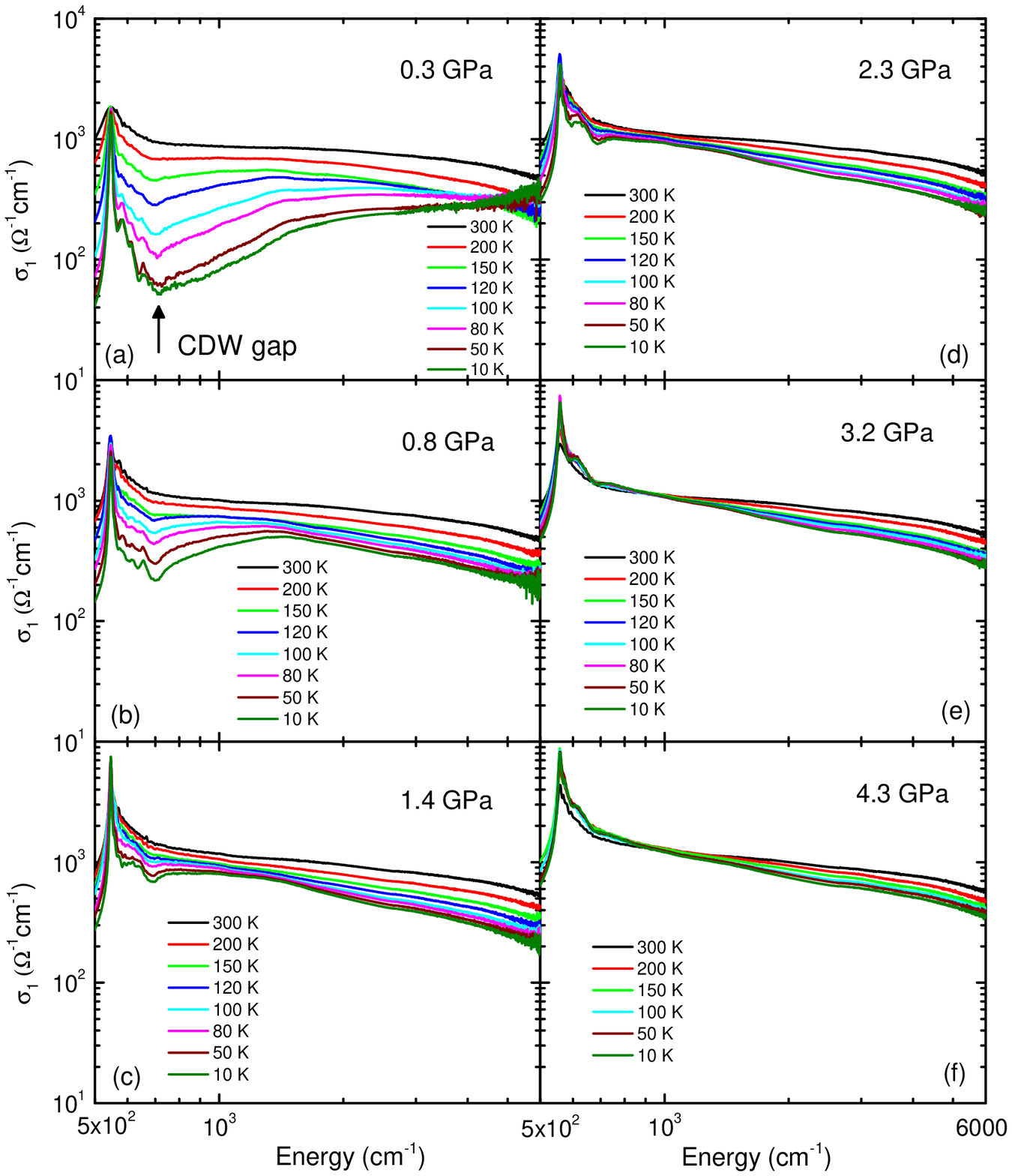}
\caption{Real part of the optical conductivity $\sigma_1$ of Sr$_{10}$Ca$_{4}$Cu$_{24}$O$_{41}$ for {\bf E}$\parallel$$c$ as a function of temperature for various pressures.}
\label{fig:cond}
\end{figure*}

\section{Experiment}
The studied Sr$_{10}$Ca$_{4}$Cu$_{24}$O$_{41}$ single crystal was grown by the floating zone
technique.\cite{Ammerahl99}
The mid-infrared reflectivity (580 - 6000 cm$^1$) at low temperature and high pressure was measured using a
home-built infrared microscope coupled to a Bruker Vertex 80/v FTIR spectrometer and maintained at the same vacuum conditions as the spectrometer, in order to avoid absorption lines of H$_2$O and CO$_2$ molecules. Thus,
the parallel beam of the radiation from the interferometer can enter the IR microscope without passing a window.
The scheme of the infrared microscope is depicted in Fig.\ \ref{fig:lthp}.
The setup can be operated either in reflection or in transmission
mode, depending on the position of the movable mirror 1. The paths
of the beam for both measuring geometries are symmetrical (see
Fig.~\ref{fig:lthp}). A spherical mirror focuses the beam on
the focal plane of the Schwarzschild objective (in reflection mode)
or on the focal plane of the Schwarzschild condenser (in transmission mode).
The infrared radiation is focused on the sample by all-reflecting home-made Schwarzschild objectives
with a large working distance of about 55 mm and 14$\times$ magnification.

A diamond anvil cell (DAC)\cite{Keller77} equipped
with type IIA diamonds suitable for infrared measurements was used to generate pressures up to 4.3~GPa. Finely
ground CsI powder was chosen as quasi-hydrostatic pressure transmitting medium.
The DAC was mounted on the cold-finger of a continuous-flow helium cryostat (Cryo\-Vac KONTI) with two metallic rods, which allow mechanical access to the DAC lever arm mechanism. Thus, the pressure in the
DAC could be changed {\it in situ} at arbitrary temperature. The pressure in
the DAC was determined {\it in situ} by the standard ruby-fluorescence technique.\cite{Mao86}
To this end, the ruby balls in the DAC are excited by a green laser with $\lambda$=532~nm, and the ruby fluorescence spectra are recorded by a grating spectrometer.
The direct observation of the sample is possible via the eyepiece or with a video camera (moving mirror 2 and 3, see
Fig.~\ref{fig:lthp}).
The above-described setup enables polarization-dependent transmission
and reflectivity measurements over a broad frequency range
(far-infrared to visible) for temperatures between $\approx$10 and 300~K,
and pressures up to 20~GPa.

Details about the geometry of
the reflectivity measurements and the Kramers-Kronig (KK) analysis of the reflectivity spectra $R_{s-d}$ at the sample-diamond interface can be found in our earlier
publications.\cite{Pashkin06,Kuntscher06}
For the KK analysis the reflectivity spectra
were fitted with the Drude-Lorentz model, in order to extrapolate the missing low- and high-frequency parts of the spectra. Hereby,
the dc resistivity data of Sr$_{11}$Ca$_{3}$Cu$_{24}$O$_{41}$ for $E$$\parallel$$c$ in Ref.\ \onlinecite{Motoyama02}
were taken into account for the low-frequency extrapolation. For the high-frequency extrapolation
the reflectivity data of Sr$_{11}$Ca$_{3}$Cu$_{24}$O$_{41}$ at room temperature\cite{Osafune97} were used for all studied temperatures and pressures. The multiphonon absorptions in the diamond anvil cause artifacts the reflectivity spectra in the frequency range 1800 - 2700~cm$^{-1}$; therefore, the spectra were interpolated in this range according to the Drude-Lorentz fitting of the reflectivity spectra.
All measurements were carried out with the polarization {\bf E} of the radiation along the crystallographic $c$ axis (ladder legs).

\begin{figure}
\includegraphics[width=0.95\columnwidth]{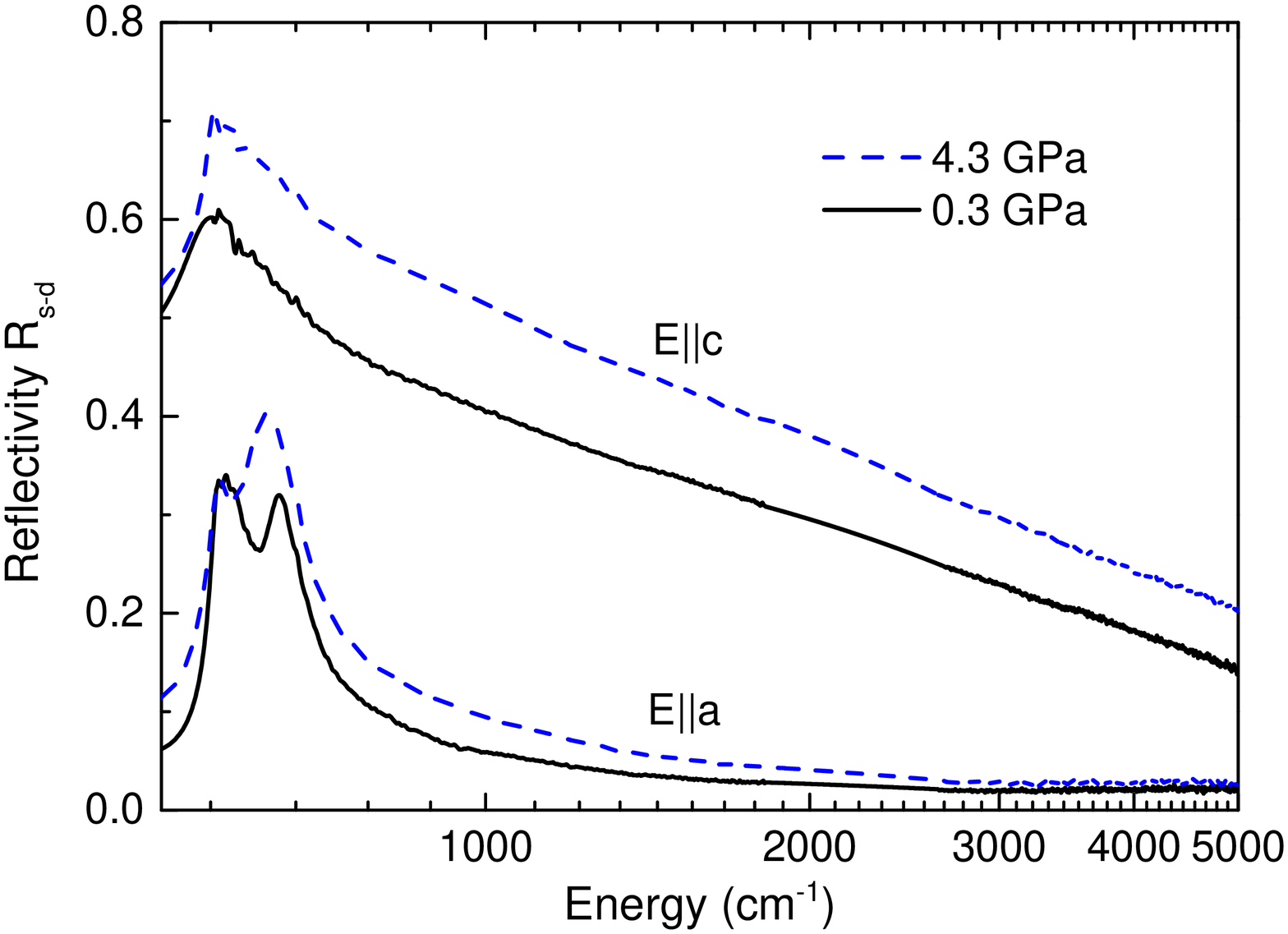}
\caption{Room-temperature reflectivity spectra of Sr$_{10}$Ca$_{4}$Cu$_{24}$O$_{41}$ for the two polarization directions {\bf E}$\parallel$$c$ and {\bf E}$\parallel$$a$ at low (0.3~GPa) and high (4.3~GPa) pressure.}
\label{fig:anisotropy}
\end{figure}

\section{Results and discussion}

The room-temperature reflectivity spectra R$_{s-d}$ of Sr$_{10}$Ca$_{4}$Cu$_{24}$O$_{41}$ at the diamond-sample interface are depicted in Fig.\ \ref{fig:room-temp} (a) for various pressures. At the lowest pressure (0.3~GPa) the reflectivity spectrum shows a plasma
edge and phonon modes at around 550~cm$^{-1}$, consistent with published data.\cite{Osafune97}
The observation of a plasma edge at optical frequencies is not in contradiction to the insulating behavior found in electrical transport measurements, since the localization of the carriers due to disorder or electronic correlations is a low-energy phenomenon.
The corresponding optical conductivity spectrum $\sigma_1$ [see Fig.\ \ref{fig:room-temp}(b)] contains considerable spectral weight in the studied frequency range, which can be attributed to excitations in the ladders.\cite{Osafune97}
With increasing pressure the plasma edge in the reflectivity spectrum shifts to higher frequencies and the
overall optical conductivity increases. In analogy to the ambient-pressure results
for Sr$_{14-x}$Ca$_x$Cu$_{24}$O$_{41}$ with increasing Ca content, we interpret this increase of spectral weight in terms of a transfer of holes from the chain to the ladder subsystem.\cite{Osafune97}

For exploring the pressure dependence of the CDW ground state, the reflectivity R$_{s-d}$ was measured for various pressures as a function of temperature down to 10~K (see Fig.\ \ref{fig:refl}). The opening of a CDW gap in the electronic band structure is manifested by an overall decrease and a change in slope of the reflectivity spectrum above 800~cm$^{-1}$, occurring at
around 150~K at the lowest pressure [see Fig.\ \ref{fig:refl}(a)]. Taking the change in slope in the reflectivity spectrum as the criterium for entering the CDW state, the CDW transition temperature $T_{CDW}$ was determined as a function of pressure: With increasing pressure $T_{CDW}$ decreases with the linear pressure coefficient of $\approx$-70~K/GPa, and at $\approx$3~GPa the CDW phase is completely suppressed. Above 3~GPa the shape of the reflectivity spectrum with a plasma edge is retained at all temperatures, and the temperature-induced changes are small. Based on the pressure and temperature evolution of the reflectivity data we propose the temperature-pressure phase diagram of Sr$_{10}$Ca$_{4}$Cu$_{24}$O$_{41}$ as shown in Fig.\ \ref{fig:pt}, including the low-temperature CDW phase and the high-temperature (HT) insulating phase.\cite{Vuletic03}

The optical conductivity spectra $\sigma_1$ as obtained by KK analysis of the reflectivity spectra are depicted in Fig.\ \ref{fig:cond}. At the lowest applied pressure (0.3~GPa), the optical conductivity decreases below 3000~cm$^{-1}$ for temperatures below 150~K, which signals the opening of the CDW gap [see Fig.\ \ref{fig:cond}(a)]. This decrease becomes more pronounced for lower temperatures.
At the lowest temperature one observes a pronounced increase in the optical conductivity above 700~cm$^{-1}$, but there is no peak structure corresponding to single-particle excitations across the CDW gap, which could serve as an energy scale for the CDW gap. We therefore use the increase in the optical conductivity above 700~cm$^{-1}$ to estimate the size of the CDW gap to $\Delta_{CDW}$$\approx$87~meV at 0.3~GPa, in analogy to Ref.\ \onlinecite{Vuletic03}.
The suppression of the CDW phase with increasing pressure is also seen in the optical conductivity spectra (see Fig.\ \ref{fig:cond}), as the temperature-induced suppression in the frequency range $<$3000~cm$^{-1}$ gradually diminishes with increasing pressure and disappears for pressures above $\approx$3~GPa. Furthermore, with increasing pressure the onset of the strong increase in the optical conductivity observed at 10~K and 0.3~GPa shifts to frequencies below the studied frequency range. Hence, the decrease of $\Delta_{CDW}$ with increasing pressure cannot be determined based on the present data.

Next we compare the effect of external pressure on the CDW state with that of increasing Ca content $x$. According to the temperature dependence of the complex dielectric response\cite{Vuletic03} the CDW transition temperature decreases with increasing $x$ and reaches the value $T_{CDW}$=20~K for $x$=9. These ambient-pressure results as a function of Ca content are included in the phase diagram in Fig. \ref{fig:pt}.\cite{comment1} From the comparison of the two data sets we obtain the simple scaling $x \approx 3\cdot P(GPa)$ between the Ca content $x$ and the applied pressure $P$.

For driving the materials Sr$_{14-x}$Ca$_x$Cu$_{24}$O$_{41}$ towards a SC state, both a high Ca content and external pressure are needed.\cite{Uehara96} Ca doping induces a more two-dimensional charge distribution in the ladder subunit according to x-ray absorption measurements.\cite{Nucker00}
It was suggested that external pressure diminishes the electronic correlations, which causes a HT insulating phase even for high Ca doping, and increases the dimensionality of the system further to two dimensional
within the $a$-$c$ ladder plane.\cite{Vuletic03} Hence, the SC state is expected to be two-dimensional in nature. A comparison of the HT reflectivity spectra for the two polarization directions {\bf E}$\parallel$$c$ (i.e., along the ladder legs) and {\bf E}$\parallel$$a$ (i.e., along the ladder rungs)
can serve as an indicator for the dimensionality of the HT insulating phase within the ladder plane. We thus depict in Fig.\ \ref{fig:anisotropy} the room-temperature reflectivity spectra of Sr$_{10}$Ca$_{4}$Cu$_{24}$O$_{41}$
for {\bf E}$\parallel$$c,a$ and for the lowest (0.3~GPa) and highest (4.3~GPa) pressure.
For the polarization direction {\bf E}$\parallel$$a$ the overall reflectivity is considerably lower as compared to {\bf E}$\parallel$$c$ and it contains two pronounced peaks at low frequencies due to phonon excitations. For an estimate of  the degree of anisotropy we use the ratio R$_{s-d}$({\bf E}$\parallel$$c$)/R$_{s-d}$({\bf E}$\parallel$$a$)
of the reflectivity spectra above 1000~cm$^{-1}$, i.e., well above the phonon excitations; the ratio amounts to about 10. With increasing pressure the overall reflectivity increases slightly for both polarization directions and the reflectivity ratio R$_{s-d}$({\bf E}$\parallel$$c$)/R$_{s-d}$({\bf E}$\parallel$$a$) is unchanged.
Obviously, the material remains highly anisotropic within the $a$-$c$ plane up to the highest applied pressure. It seems that the dimensionality of the HT insulating phase is only slightly affected by
external pressure for such a low Ca content.
Also for high Ca contents the anisotropic character of the HT insulating
state is preserved under external pressure, as demonstrated by the pressure dependence of the resistivity ratio
$\rho_a$/$\rho_c$ for Sr$_{2.5}$Ca$_{11.5}$Cu$_{24}$O$_{41}$ at room temperature\cite{Nagata98}
and by the pressure-dependent lattice constants in Sr$_{0.4}$Ca$_{13.6}$Cu$_{24}$O$_{41}$.\cite{Pachot05}
The robustness of the anisotropic nature of the HT insulating state against pressure thus seems to be common to all
compounds Sr$_{14-x}$Ca$_x$Cu$_{24}$O$_{41}$.

\section{Conclusions}

In conclusion, from our pressure- and temperature-dependent optical reflectivity data on Sr$_{10}$Ca$_{4}$Cu$_{24}$O$_{41}$ we infer a linear decrease of the CDW transition temperature $T_{CDW}$, with a linear pressure coefficient \linebreak $\approx$-70~K/GPa. Above $\approx$3~GPa the CDW phase is completely suppressed. By comparing the pressure dependence of $T_{CDW}$ with the ambient-pressure values of $T_{CDW}$ in the compounds Sr$_{14-x}$Ca$_x$Cu$_{24}$O$_{41}$ for various Ca contents $x$, we find the simple scaling $x \approx 3\cdot P(GPa)$ between $x$ and the applied pressure $P$. The size of the CDW gap decreases with increasing pressure.
The HT insulating phase in Sr$_{10}$Ca$_{4}$Cu$_{24}$O$_{41}$ preserves its anisotropy within the ladder plane up to the highest applied pressure.

\section{Acknowledgment}

We thank K. Syassen for providing valuable information about the optical design of the home-made
infrared microscope and A. Pashkin for the optical design and construction of the home-made infrared microscope.
This work is financially supported by the DFG (KU 1432/6-1).

{}


\begin{thebibliography}{}
\bibitem{comment1}
Please note that in Refs.\ \onlinecite{Rusydi06,Osafune97a} a different dependence of the CDW on the Ca content was found.
\end{thebibliography}
\end{document}